\documentclass[aps,prd,10pt,twocolumn,preprintnumbers,superscriptaddress,bibnotes,nofootinbib,letterpaper,longbibliography]{revtex4-2}
\usepackage[english]{babel}

\usepackage{amsfonts}
\usepackage{amsmath,mathtools,physics,xfrac}
\usepackage[normalem]{ulem}
\usepackage{graphicx}
\usepackage{afterpage}
\usepackage{float}
\usepackage{subfigure}
\usepackage{rotating}
\usepackage{multirow}
\usepackage{tabularx}
\usepackage{booktabs}
\usepackage{fancyhdr}
\usepackage[utf8]{inputenc}
\usepackage{theorem}
\usepackage{moreverb}
\usepackage{euscript}
\usepackage{psfrag}
\usepackage{slashed}
\usepackage{mathtools}
\usepackage{makecell}
\usepackage{adjustbox}
\usepackage{dcolumn}
\usepackage{bm}
\usepackage[dvipsnames]{xcolor}
\usepackage{graphics}
\usepackage{hyperref}
\usepackage{lettrine}

\input Zallman.fd

\definecolor{darkviolet}{rgb}{0.58, 0.0, 0.83}

\hypersetup{
     colorlinks   = true,
     citecolor    = darkviolet,
     urlcolor     = darkviolet,
     linkcolor    = darkviolet
}

\usepackage{color,xcolor}

\definecolor{darkblue}{rgb}{0.0,0.0,0.75}
\definecolor{darkred}{rgb}{0.6,0.0,0}
\definecolor{darkgreen}{rgb}{0.0,0.6,0.}

\newcommand\redsout{\bgroup\markoverwith{\textcolor{red}{\rule[0.5ex]{2pt}{0.4pt}}}\ULon}

\begin{document}

\preprint{SLAC-PUB-261007}

\title{Revealing Dark Matter in the Dark-Electron Rate of Xenon Detectors}

\author{Lillian Santos-Olmsted}
\thanks{\href{mailto:solmsted@stanford.edu}{solmsted@stanford.edu}, \href{https://orcid.org/0000-0002-8763-3702}{0000-0002-8763-3702}}
\affiliation{Particle Theory Group, SLAC National Accelerator Laboratory, Stanford, CA 94305, USA}
\affiliation{Kavli Institute for Particle Astrophysics and Cosmology, Stanford University, Stanford, CA 94305, USA}

\author{Rebecca~K.~Leane}
\thanks{\href{mailto:rleane@stanford.edu}{rleane@stanford.edu}, \href{https://orcid.org/0000-0002-1287-8780}{0000-0002-1287-8780}}
\affiliation{Particle Theory Group, SLAC National Accelerator Laboratory, Stanford, CA 94305, USA}
\affiliation{Kavli Institute for Particle Astrophysics and Cosmology, Stanford University, Stanford, CA 94305, USA}

\author{Carlos Blanco}
\thanks{\href{mailto:carlosblanco@psu.edu}{carlosblanco@psu.edu}, \href{https://orcid.org/0000-0001-8971-834X}{0000-0001-8971-834X}}
\affiliation{Institute for Gravitation and the Cosmos, The Pennsylvania State University, University Park, PA 16802, USA}
\affiliation{Department of Physics, Princeton University, Princeton, NJ 08544, USA}
\affiliation{Stockholm University and The Oskar Klein Centre for Cosmoparticle Physics, Alba Nova, 10691 Stockholm, Sweden}

\author{John F. Beacom}
\thanks{\href{mailto:beacom.7@osu.edu}{beacom.7@osu.edu}, \href{https://orcid.org/0000-0002-0005-2631}{0000-0002-0005-2631}}
\affiliation{Center for Cosmology and AstroParticle Physics (CCAPP),
Ohio State University, Columbus, Ohio 43210, USA}
\affiliation{Department of Physics, Ohio State University, Columbus, Ohio 43210, USA}
\affiliation{Department of Astronomy, Ohio State University, Columbus, Ohio 43210, USA}

\date{\today}

\begin{abstract}

There is high interest in improving experimental sensitivity to sub-GeV dark matter (DM) that scatters with electrons, as compelling theoretical models seem within reach. At the lowest DM masses, most of the expected signal lies in the single-electron regime, but so do large backgrounds. Xenon experiments therefore either do not yet search there at all or, as in XENONnT, apply cuts so severe that only $\sim$$10^{-4}$ of the exposure is retained, with sensitivity set by the criterion $S \lesssim B$, where those are the signal and background counts. We advocate that xenon detectors instead utilize most of their dark-electron rate (i.e., the total raw electron hit rate) and perform an annual-modulation search, for which the sensitivity scales as $S \lesssim \sqrt{B}$. Modern DAQs already record this rate, continuously in triggerless systems such as the XENONnT and PandaX-4T DAQs, and within trigger windows in LZ, so no new hardware is required. Using measured dark-electron rates, we estimate the sensitivity of XENONnT, LZ, PandaX-4T, PandaX-xT, and XLZD. We find that dramatic gains in sensitivity are possible, motivating new analyses of present and future xenon data.
\end{abstract}

\maketitle


\section{Introduction}

\lettrine{T}{he search for sub-gev dark matter (dm)} is increasingly limited not by the
ability of detectors to register low-energy quanta, but by the challenge of
distinguishing signals from detector backgrounds. This is particularly true for DM-electron
scattering~\cite{Essig:2011nj, Essig:2012yx, Graham:2012su, Essig:2015cda, Hochberg:2015pha, Hochberg:2015fth, Derenzo:2016fse, Hochberg:2016ntt, Essig:2017kqs, Hochberg:2017wce, Cavoto:2017otc, Emken:2017erx, Sanchez-Martinez:2019bac, Hochberg:2019cyy,  Essig:2019xkx, Emken:2019tni, Kurinsky:2019pgb, Geilhufe:2019ndy, Blanco:2019lrf, Baxter:2019pnz, Catena:2019gfa, Radick:2020qip, Gelmini:2020xir, Trickle:2020oki, Kurinsky:2020dpb, Griffin:2020lgd, Blanco:2021hlm, Knapen:2021run, Hochberg:2021ymx, Hochberg:2021yud, Hochberg:2022apz, Blanco:2022cel, Cook:2024cgm, QROCODILE:2024nqm, Griffin:2024jec,Leane:2025efj,Santos-Olmsted:2025nuk,Griffin:2025wew,Marocco:2025eqw,Hochberg:2025dom,Hochberg:2025rjs,Hochberg:2026mdl,Kozynets:2026psw}, where light DM can sufficiently ionize target materials, producing electrons that await new techniques to efficiently identify them. Dual-phase xenon time projection chambers (TPCs), such as LUX-ZEPLIN (LZ)~\cite{LZ:2019sgr, LZ:2022lsv, LUX-ZEPLIN:2025uaf}, XENONnT~\cite{XENON:2022ltv, XENON:2024wpa, XENON:2024znc}, and PandaX-4T~\cite{PandaX-4T:2021bab, PandaX:2022xqx, PandaX:2024med}, are particularly ripe for investigating this regime. This is because they combine multi-ton target masses with large electroluminescence gain for extracted electrons, thereby enabling sensitivity to very small ionization signals.

\begin{figure}[t!]
  \centering
  \includegraphics[width=0.96\columnwidth]{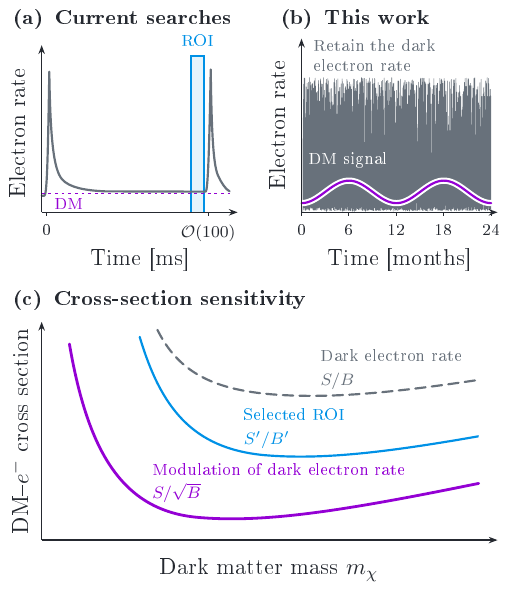}
  \caption{Search strategies for single-electron sensitivity.
  \textbf{(a)} Current searches cut on time after large S2 depositions, selecting a region of interest (ROI) with improved signal-to-background ratio $S'/B'$, at a large cost to exposure.
  \textbf{(b)} This work: the DM annual modulation is separated statistically while retaining most of the exposure, so the sensitivity scales as $S/\sqrt{B}$. \textbf{(c)} Qualitative comparison of the resulting sensitivities.}
  \label{fig:strategy}
\end{figure}

Accessing this low-energy regime requires going below the thresholds of full event reconstruction, which conventionally relies on both prompt scintillation (S1) and ionization (S2) signals. While positive scintillation identification requires energy depositions in the keV range, S2-only analyses, which focus only on the ionization signal, can be sensitive to lower recoil energies~\cite{Essig:2011nj, Essig:2012yx, Essig:2017kqs, XENON:2019gfn, XENON:2021qze}. These analyses still typically impose thresholds of $\sim$100--150 detected photoelectrons ($\sim$3--5 extracted electrons), for an interaction to be classified as an event~\cite{XENON:2019gfn, XENON:2022ltv, LUX-ZEPLIN:2025uaf}. 

Thresholds can be reduced even further by using the single-electron rate~\cite{Essig:2011nj, XENON:2021qze, XENON:2024znc}, where most of the signal from the lightest DM is expected. In single-electron analyses, the threshold is lowered to the xenon ionization energy, ${\cal O}(10\,{\rm eV})$. However, it is much harder to discriminate signal from background in the single-electron regime. Experiments therefore either do not yet search there at all, or do so only with severe cuts. XENON1T and XENONnT are the only xenon experiments to have performed a dedicated search for sub-GeV DM down to the single-electron regime~\cite{XENON:2021qze, XENON:2024znc}, using temporal and spatial selections that, in the lowest-electron bins, reduce the usable exposure by many orders of magnitude. 

In this paper, we point out a way to push the light DM sensitivity to even smaller cross sections. Rather than aggressively cutting backgrounds, we propose to use the raw electron hit rate, which we call the \textit{dark-electron rate} in analogy to the photomultiplier tube (PMT) dark rate and the dark current of skipper CCDs, and to search it for the annual modulation induced by Earth's motion through the Galactic DM halo~\cite{Drukier:1986tm}. Annual modulation has been searched for in keV-scale xenon data~\cite{XENON100:2015hgt, XENON:2017nik, LUX:2018xvj, XMASS:2018koa} using reconstructed events, and proposed for post-cut ionization data in electron-number bins~\cite{Essig:2017kqs}. \textit{Our strategy is qualitatively different. We consider the time series of most of the recorded electron hits, $\mathit{\sim\!10^{9}}$ per year, independent of event-level classification and with only minimal cuts.} Retaining these hits pays off because modulation sensitivity scales as $S/\sqrt{B}$ rather than $S/B$, so a cut that keeps a fraction $\epsilon$ of the exposure improves the reach only if it keeps less than $\epsilon^2$ of the background counts. The single-electron selections of XENONnT discussed above, designed for a search limited by $S/B$, reduce the exposure by $\sim$$10^{4}$ but the background rate by only $\sim$$10^{2}$~\cite{XENON:2024znc, Peres:2023rdv, Riya:2025eze}. Figure~\ref{fig:strategy} contrasts the two approaches.

Our strategy would have several advantages. First, it retains orders of magnitude more exposure, as individual electrons need not pass the stringent cuts used to define a low-background sample. Second, no new hardware is needed. The triggerless data acquisition systems (DAQs) of XENONnT and PandaX-4T record the dark-electron rate throughout essentially the full runtime~\cite{XENON:2022vye, Yang:2021hnn}, providing an enormous, currently unused sample. LZ records the rate within its trigger windows~\cite{LZ:2022lsv, LZ:2024bvw, LUX-ZEPLIN:2025uaf}, so it can also search now, though a triggerless upgrade would increase its exposure. Third, the cryogenic, highly stable environment of dual-phase TPCs limits environmental effects that could mimic a modulation. Finally, several independent experiments at different underground sites can perform the search, and a DM signal should appear in all of them with a common phase and predictable relative amplitudes, serving as a cross-check against detector-specific systematics.


\section{DM Signal in Xenon TPCs}
An energy deposition in a dual-phase xenon TPC produces prompt scintillation (S1) and ionization electrons, which drift upward and are extracted into the gas phase, where each produces a secondary scintillation signal (S2) of $\sim$$800$ photons via electroluminescence (leading to $\sim$$30$ detected photoelectrons in the top PMT array). This large gain makes even single extracted electrons detectable, so the DM observable is the rate of extracted electrons. Below, we compute the DM-induced electron rate and then describe the detector inputs that set the detection efficiency and exposure.

\begin{table*}[t]
\centering
\small
\setlength{\tabcolsep}{7pt}
\begin{tabular}{lccccccc l}
\hline\hline
 & $M_{\rm act}$ [ton] & Live time [\%] & $t_d$ [ms] & $\tau_e$ [ms] & $P_I$ [\%] & $P_E$ [\%] & $\xi$ & References \\
\hline
LZ & 7.0 & $\sim5$ & 0.8--1 & 5--14 & 91--97 & 72--98 & 0.57--0.95 &
\cite{Mount:2017qzi, LZ:2019sgr, LZ:2024zvo, LUX-ZEPLIN:2025uaf, LZ:2022lsv, LZ:2024bvw} \\

XENONnT & 5.9 & $ \gtrsim 99$ & 2.2 & 5--20 & 81--95 & 50--56 & 0.35--0.53 &
\cite{XENON:2024mlv,XENON:2024kbh,Kopec:2023uii,XENON:2022ltv,XENON:2024wpa,XENON:2024znc, XENON:2022vye} \\

PandaX-4T & 3.7 & $ \gtrsim 99$ & 0.8--0.85 & 0.6--1.7 & 53--80 & 85--98 & 0.39--0.78 &
\cite{PandaX-4T:2021bab, PandaX:2024med, Yang:2021hnn, PandaX:2024cic, He:2021sbc} \\

PandaX-xT & 43 & $ \gtrsim 99$ & -- & $\gtrsim 2$ & 53--80 & 85--98 & 0.39--0.78 & 
\cite{PandaX-4T:2021bab, PandaX:2024med, Guo:2026frk, PANDA-X:2024dlo} \\

XLZD & 60 & $ \gtrsim 99$ & -- & $\gtrsim 10$ & 81--97 & 85--98 & 0.60--0.95 & 
\cite{XLZD:2024nsu, XLZD2025ESPPU} \\
\hline\hline
\end{tabular}
\caption{Detector parameters and resulting detection efficiency $\xi$, using $P_R \simeq 87$--$100\%$~\cite{XENON:2019izt, Szydagis:2011tk, XENON:2021qze, Szydagis:2022ikv}. $M_{\rm act}$ denotes the total active target mass in metric tons. For PandaX-xT, we use the PandaX-4T efficiency values.}
\label{tab:efficiency}
\end{table*}

\subsection{DM Signal Calculation}

A DM-induced ionization yields a primary electron, plus secondary electrons from the ionization of other atoms~\cite{Essig:2012yx, Essig:2017kqs}, so the total detected rate is
\begin{equation}
    R_{\text{tot}} = R_{\text{prim}} + R_{\text{sec}}. 
\end{equation}
Primary electrons are detected at a rate $R_{\text{prim}}$,
\begin{equation}
\label{eq:Rprim}
    R_{\text{prim}} = \xi \int \frac{dR}{d\ln E_{er}} \frac{dE_{er}}{E_{er}},
\end{equation}
where $E_{er}$ is the energy of the ejected electron after ionization, $dR/{d\ln E_{er}}$ is the differential ionization rate, and $\xi$ is the detector's efficiency factor, which is discussed in the next subsection. The differential ionization rate is~\cite{Essig:2011nj}
\begin{align}
\label{eq:diffRate}
    &\frac{dR}{d\ln E_{er}} = \frac{\rho_{\chi} \bar{\sigma}_e}{8 m_{\chi} \mu_{e \chi}^2} N_T   \\ &\times \sum_i \int |f_{\rm{ion}}^{i}(E_{er},q)|^2 \ |F_{\text{DM}}(q)|^2 \ \eta(v_{\text{min}}) \ q \ dq,\nonumber
\end{align}
where $\rho_{\chi} = 0.4 \ \rm{GeV}/cm^3$ is the local DM density, $m_{\chi}$ is the DM mass, $\mu_{e \chi}$ is the DM-electron reduced mass, and the number of targets $N_T$ is the number of xenon atoms in the detector. Inside the integral over the momentum transfer $q$, the ionization form factor $|f_{\rm{ion}}^{i}(E_{er},q)|^2$ encodes the probability of ionization for an initial bound state $i$ corresponding to an atomic orbital. We use xenon form factors from DarkART~\cite{Catena:2019gfa, DarkART}. The DM form factor $|F_{\text{DM}}(q)|^2$ incorporates the momentum dependence of the free electron-DM scattering amplitude, and $\bar{\sigma}_e$ is the DM-electron cross section evaluated at $q = \alpha m_e$, where $\alpha$ is the fine structure constant and $m_e$ is the electron mass. The mean inverse velocity of the DM is 
\begin{equation}
    \eta(v_{\text{min}}) = \int{d^3v \ g_\chi(\vec{v})\frac{1}{v}\Theta(v-v_{\text{min}})},
\end{equation}
where 
\begin{equation}
    v_{\text{min}} = {(|E_b^i|+E_{er})}/{q}+{q}/{2 m_{\chi}}
\end{equation}
is the minimum DM velocity required to transfer energy $|E_b^i|+E_{er}$ to the electron, and $E_b^i$ is the binding energy. For the DM velocity distribution $g_\chi(\vec{v})$, we adopt a spherically symmetric Maxwell-Boltzmann distribution with local mean velocity $v_0 = 230 \ \rm{km}/s$, Galactic escape velocity $v_{\rm esc}=600 \ \rm{km}/s$, and Earth velocity in the Milky Way rest frame $v_E = 240 \ \rm{km}/s$~\cite{Essig:2015cda}. 

Secondary ionization of other atoms can be induced by high-energy primary electrons or by de-excitation photons generated when inner orbitals are ionized and an outer-shell electron fills that vacancy~\cite{Essig:2012yx, Essig:2017kqs}. Following Ref.~\cite{Essig:2017kqs}, the number of quanta created from high-energy primary electrons is $n^{(1)} = \text{Floor}(E_{er}/W)$, where $W$ is the energy required to produce a quantum (exciton or electron) in liquid xenon~\cite{XENON:2019izt}. The contribution from de-excitation photons $n^{(2)}$ depends on the original ionized orbital. For the $5p$ and $5s$ orbitals, $n^{(2)} = 0$. The number of quanta produced from the ionization of the ($4d$, $4p$, $4s$) orbitals is $n^{(2)}$=(4, 6--10, 3--15), where the ranges account for the fact that the outer-shell electron can come from one of multiple outer orbitals. As in Ref.~\cite{Essig:2017kqs}, we select the lowest number in the range to remain conservative. The probability that each quantum produces an electron that survives recombination is~\cite{XENON:2021qze}
\begin{equation}
    p_e = (1 - \langle r \rangle)/(1+\langle N_{\text{ex}}/N_{\text{ion}}\rangle)\,,
\end{equation}
where $\langle r \rangle$ is the average recombination fraction and $\langle N_{\text{ex}}/N_{\text{ion}}\rangle$ is the exciton-to-ion ratio. The average number of secondary electrons produced per ionization is $p_e(n^{(1)}+n^{(2)})$. Therefore, combining this average number with Eq.~(\ref{eq:Rprim}) and Eq.~(\ref{eq:diffRate}), the rate of detected secondary electrons is 
\begin{align}
\label{eq:Rsec}
    &R_{\text{sec}} = \xi \int \frac{dE_{er}}{E_{er}} \frac{\rho_{\chi} \bar{\sigma}_e}{8 m_{\chi} \mu_{e \chi}^2} N_T \sum_i \frac{n^{(1)}+n^{(2)}}{1+\langle N_{\text{ex}}/N_{\text{ion}}\rangle}  \\ &\times  \int |f_{\rm{ion}}^{i}(E_{er},q)|^2 \ |F_{\text{DM}}(q)|^2 \ \eta(v_{\text{min}}) \ q \ dq\nonumber\,, 
\end{align}
where the factor of $1 - \langle r \rangle$ in $p_e$ was absorbed in the efficiency factor $\xi$, which we discuss below. 


\begin{figure*}[t]
    \centering
    \includegraphics[width=\textwidth]{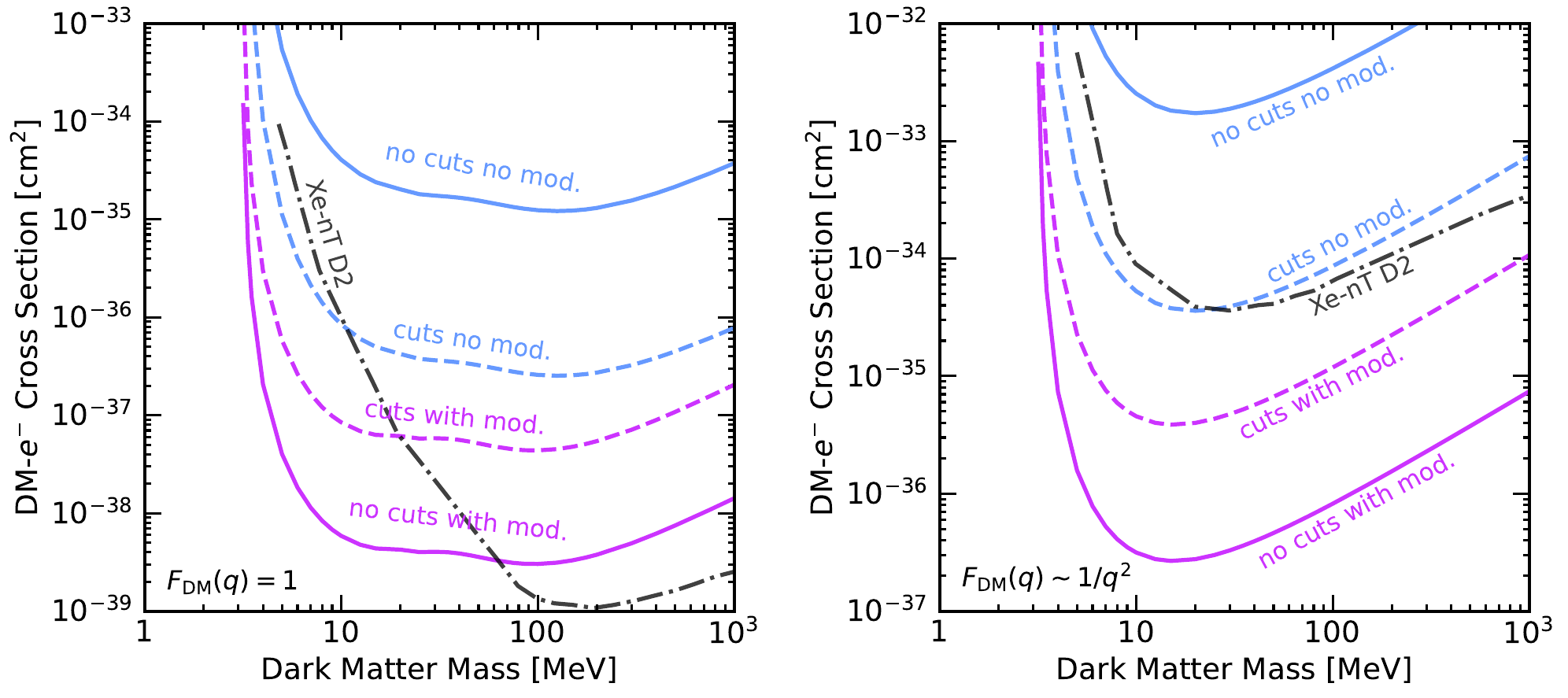}
    \caption{Sensitivity of XENONnT for a heavy mediator (left panel) and light mediator (right panel); note different y-axis ranges.  We calculate and compare four cases: the uncut dark-electron rate without modulation (solid blue), the cut rate without modulation (dashed blue), the cut rate with annual modulation (dashed pink), and the uncut rate with modulation (solid pink). The black dot-dashed line is the S2-only limit from the XENONnT D2 dataset, which includes multi- and single-electron rates and no modulation~\cite{XENON:2024znc}. For all lines, we assume an average detection efficiency of $\xi \simeq 0.44$ and an exposure of~$5.9$~$\text{ton}\times\text{yrs}$.}
    \label{fig:sensitivity_comparison}
\end{figure*}

\subsection{Detector Parameters}
The xenon TPCs we consider (LZ, XENONnT, PandaX-4T, PandaX-xT, and XLZD) differ in target mass, live time, and detection efficiency. Table~\ref{tab:efficiency} summarizes the representative values we adopt. The target mass sets $N_T$ in Eq.~\eqref{eq:diffRate}, and the live time is the fraction of runtime over which the dark-electron rate is recorded, which is $\sim$5\% for LZ's current triggered DAQ, which records 4.5~ms windows around $\sim5\,$Hz of event triggers plus $1\,$Hz GPS and $4\,$Hz random triggers~\cite{LZ:2022lsv, LZ:2024bvw, LUX-ZEPLIN:2025uaf}, and $\gtrsim$99\% for the triggerless DAQs of the others, which store essentially every PMT hit~\cite{XENON:2022vye, He:2021sbc}. For a uniform comparison, and because LZ could adopt a triggerless DAQ, we assume full DAQ live time for all detectors. The remaining input is the probability of detecting an ionized electron,
\begin{equation}
    \xi = P_R \times P_I \times P_E,
\end{equation}
where each factor corresponds to a stage of the electron's path to detection: surviving recombination at the interaction site ($P_R$), surviving capture by impurities while drifting ($P_I$), and being extracted into the gas phase ($P_E$).

First, an ionized electron can recombine with a xenon ion before it begins to drift. The survival probability is~\cite{XENON:2021qze}
\begin{equation}
    P_R = 1 - \langle r \rangle = W Q_y\left(1 + \langle N_{\text{ex}}/N_{\text{ion}}\rangle\right),
\end{equation}
where $W = 13.7 \pm 0.2$~eV~\cite{XENON:2019izt} and $Q_y$ is the charge yield. For $\langle N_{\text{ex}}/N_{\text{ion}}\rangle \simeq 0.06$--$0.2$~\cite{Szydagis:2011tk, Szydagis:2022ikv} and an approximately constant $Q_y \simeq 60$--$80~e^-/\text{keV}$ at low energies ($E_{er} \lesssim 200$~eV), depending on the recombination model~\cite{XENON:2021qze}, this gives $P_R \simeq 87$--$100\%$ for all detectors.

Second, as the electron drifts upward, it can be captured by electronegative impurities such as O$_2$, surviving with probability $e^{-t/\tau_e}$ after drift time $t$ for electron lifetime $\tau_e$~\cite{Kopec:2023uii,XENON:2024mlv}. Averaging over the detector height, with maximum drift time $t_d$,
\begin{equation}
    P_I \simeq \frac{1}{t_d} \int_0^{t_d} e^{-t/\tau_e}\, dt = \frac{\tau_e}{t_d} \left(1-e^{-t_d/\tau_e}\right).
\end{equation}
For XLZD, $t_d$ is not yet fixed and $\tau_e \gtrsim 10$~ms is expected~\cite{XLZD:2024nsu}, so we adopt the combined range of $P_I$ from LZ and XENONnT. 

Third, on reaching the liquid surface, the electron must cross into the gas phase, where it produces electroluminescence. The extraction efficiency $P_E$ increases with the extraction field~\cite{Xu:2019dqb}, so it is lower in XENONnT than in LZ~\cite{XENON:2024wpa,Mount:2017qzi,Linehan:2021qnb}. XLZD aims for full extraction~\cite{XLZD2025ESPPU}, but we use the best $P_E$ range of the current detectors (LZ, XENONnT, PandaX-4T) to represent an optimistic goal of what could be achieved in XLZD. For PandaX-xT, whose parameters are not yet known, we adopt the PandaX-4T values for all three factors.


\begin{figure*}[t]
    \centering
    \includegraphics[width=\textwidth]{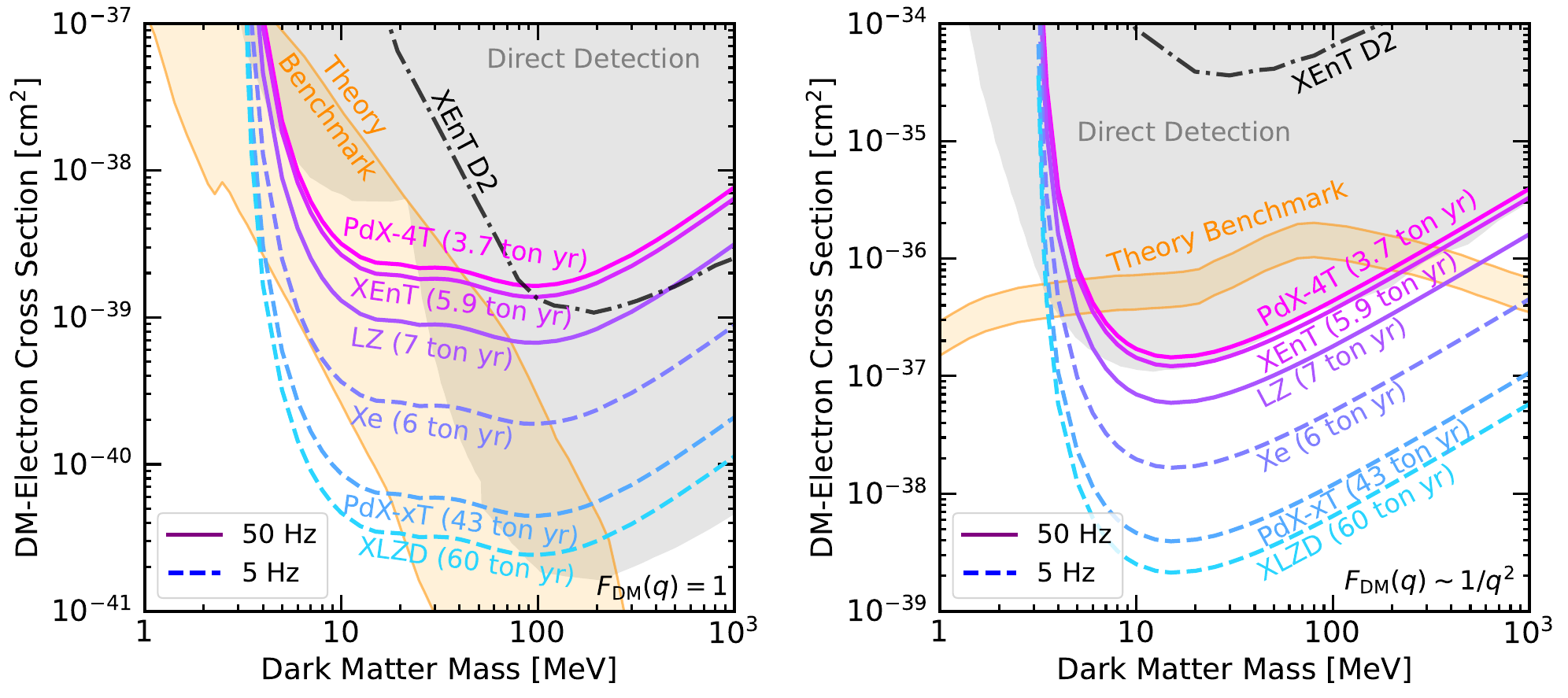}
    \caption{Projected 95\% C.L. sensitivity to DM-electron scattering cross sections for current and upcoming dual-phase xenon experiments using the dark-electron rate with annual modulation for a heavy mediator (left panel) and light mediator (right panel). These sensitivities assume one year of exposure with the active xenon mass of each detector, and 100\% live time (though LZ's current DAQ has $\sim$5\% live time). The solid lines represent sensitivities using the current measured dark-electron rate of approximately $50$~Hz after minimal cuts~\cite{LZ_WS2025_LowMassWIMP}, while the dashed lines assume an optimistic future rate of $5$~Hz. Current direct-detection limits are shown as the shaded region~\cite{DAMIC:2019dcn, XENON:2019gfn, SuperCDMS:2020ymb, SENSEI:2020dpa, PandaX-II:2021nsg, DarkSide:2022knj, DAMIC-M:2025luv, Cheek:2025nul, PandaX:2025rrz}, while the black dot-dashed line represents the ionization-only limits from XENONnT's D2 dataset~\cite{XENON:2024znc}, extending down to the single-electron regime. We also show benchmark theory model space~\cite{Essig:2022dfa}. Note the different y-axis ranges.}
    \label{fig:sensitivity}
\end{figure*}

\section{DM Signal Extraction and Sensitivity}
The dark-electron rate incorporates the full stream of raw electron hits, which includes any DM signal as well as the background. We do not separate the hits into electron-number bins, but the dark-electron rate is dominated by the single-electron background. Single-electron backgrounds are an active area of investigation~\cite{XENON:2021qze, XENON:2024znc, LUX-ZEPLIN:2025uaf} and are expected to be dominated by delayed electron emission: electrons from larger S2 depositions are captured by impurities and later released, arriving after the maximum drift time has passed. The single-electron rate therefore falls with time after each large S2, eventually as a power law~\cite{XENON:2021qze, LUX-ZEPLIN:2025uaf}. It is also enhanced near the detector walls, around the electrode wires, and spatially close to the most recent large S2 events~\cite{XENON:2021qze, XENON:2024znc, LUX-ZEPLIN:2025uaf}. Current DM searches in the single-electron regime therefore retain only time intervals long after large S2 depositions and remove these spatial regions, at a large cost in exposure.

We instead seek to distinguish signal from background only statistically, using the annual modulation of the DM velocity relative to Earth. The modulation fraction is
\begin{equation}
    f_{\text{mod}} = \frac{R_{\text{high}}-R_{\text{low}}}{2 R_0},
\end{equation}
where $R_0$ is the rate calculated using the standard value of $v_E$, $R_{\rm high}$ is the rate obtained when adding 15\,km/s to $v_E$, and $R_{\rm low}$ is the rate when subtracting 15\,km/s from $v_E$. We set the 95\% C.L. sensitivity by $f_{\rm mod} S = 2\sqrt{B}$~\cite{Essig:2017kqs} (where we assume $S \ll B$), with number of signal counts $S$ and background counts $B$ over the exposure time, taking $B$ to be the time-averaged dark-electron rate of a given experiment. This relation assumes a statistics-limited regime, as discussed below in Sec.~\ref{sec:feasible}, and follows from the fact that the modulating part of the signal ($f_{\text{mod}}S$) should not exceed the background uncertainty ($\sqrt{B}$) in the statistics-limited regime, with the factor of $2$ corresponding to the 95\% C.L. sensitivity. A full experimental analysis would instead fit the measured time-dependent rate directly. 

\subsection{Retaining Exposure vs.\ Cutting Background}
\label{sec:cuts}

Figure~\ref{fig:sensitivity_comparison} compares analysis strategies for XENONnT, which we use as an example because it has an existing single-electron analysis~\cite{XENON:2024znc}. The colored lines are our simplified estimates, which isolate the effect of the analysis strategy rather than reproduce a full experimental analysis. To produce our estimates, we use the measured uncut dark-electron rate published by XENONnT, which is $\sim$$200$~Hz~\cite{Peres:2023rdv, Riya:2025eze}, as well as the measured rate after extensive cuts (schematically matching the ROI shown in Fig.~\ref{fig:strategy}), $\sim$$1$~Hz~\cite{XENON:2024znc}, normalized to the reduced fiducial mass, and where the background and signal counts scale with the same fraction of retained live time. This ROI selection reduces the exposure in the single-electron bin by $\sim$$10^4$~\cite{XENON:2024znc}, mostly through temporal cuts, with volume cuts contributing at most a reduction factor of $\sim$$4$. We show sensitivities corresponding to both the cut and uncut datasets, with (using $f_{\rm mod} S/\sqrt{B}$) and without (using $S < B$) modulation.  For comparison, the black dot-dashed line shows the published XENONnT result from its ionization-only D2 dataset (the second of the two datasets in Ref.~\cite{XENON:2024znc}), which is a full experimental analysis without modulation that includes all selections, efficiencies, and electron-number binning.

Figure~\ref{fig:sensitivity_comparison} shows that different analysis strategies may be optimal depending on whether annual modulation is included in the analysis. Without modulation, the cut dark-electron rate gives stronger sensitivity, as its smaller background yields a larger signal-to-noise ratio, justifying the traditional approach. With annual modulation, however, the sensitivity scales as $f_{\rm mod} S/\sqrt{B}$, so the $10^4$ loss in exposure outweighs the $\sim$$10^2$ reduction in background rate, and the uncut dark-electron rate gives the stronger sensitivity. In a modulation search, it is statistically speaking therefore favorable to retain the bulk of the exposure, despite the larger background.

The published XENONnT result reflects the cost of the exposure cuts. Our simplified XENONnT cut estimate without modulation (blue dashed), physically corresponding to the XENONnT analysis strategy, roughly reproduces their result wherever the published sensitivity is driven by the lowest electron-number bins. For a heavy mediator, this holds at low masses, where the signal falls mostly in the single-electron bin. At higher masses, the published result is stronger, as it draws on higher electron-number bins with smaller backgrounds and less severe cuts. For a light mediator, the $1/q^2$ form factor favors low-energy recoils at all masses, so the published result is dominated by the lowest bins throughout, and our estimate agrees roughly across the full mass range. This agreement supports using our simplified treatment to compare strategies, and shows that the sharp loss of sensitivity at low masses is driven by the exposure cuts in the lowest bins.

\subsection{Projected Sensitivities}

Figure~\ref{fig:sensitivity} shows our estimated sensitivities for PandaX-4T, XENONnT, LZ, PandaX-xT, and XLZD using the dark-electron rate with annual modulation. We assume one year of runtime with the full active mass and continuous recording of the dark-electron rate, and the average of the minimum and maximum detection efficiencies in Table~\ref{tab:efficiency}. For LZ, XENONnT, and PandaX-4T, we adopt the dark-electron rate measured by LZ after minimal cuts, $\sim$$50$~Hz~\cite{LZ_WS2025_LowMassWIMP}, which removes only $\sim$10\% of the live time and reduces the background rate by a factor of $\sim$$4$ from its uncut value. The XENONnT dark-electron rates used in Sec.~\ref{sec:cuts} are either fully uncut or severely cut, and no equivalent minimally cut rate has been published for XENONnT or PandaX-4T, so for simple direct comparison we use the LZ value for all three detectors. Our estimates improve on the XENONnT D2 sensitivity~\cite{XENON:2024znc} by one to two orders of magnitude at low masses, and reach the parameter space of benchmark models that produce the observed relic abundance (thermal freeze-out for the heavy mediator, freeze-in for the light mediator)~\cite{Essig:2022dfa}.

These estimates assume 100\% DAQ live time for a uniform comparison. LZ's current DAQ, however, records the dark-electron rate only in windows around triggers, corresponding to $\sim$5\% of the runtime~\cite{LZ:2022lsv, LZ:2024bvw, LUX-ZEPLIN:2025uaf}. LZ can perform this analysis now with that exposure, but our results strongly motivate a triggerless DAQ, as in XENONnT and PandaX-4T. This would carry little storage cost, as single-electron S2 depositions were only $\sim$2\% of the total data volume in XENONnT's first science run~\cite{XENON:2022vye}.

We also show the potential reach of xenon detectors in the future assuming that the dark-electron rate can be reduced to $\sim$$5$~Hz, indicated by the dashed lines. This optimistic benchmark could be achieved through improved suppression of delayed emission, for example by reducing electron trapping, without significantly reducing the usable exposure. For this rate, we show a 6-ton xenon detector with $\xi = 1$, representative of the target mass of current TPCs, alongside upcoming experiments PandaX-xT and XLZD. These aim to scale up the DAQ systems of current detectors~\cite{He:2021sbc, Guo:2026frk, XLZD:2024nsu}. Note that the upgrade from PandaX-4T to PandaX-xT will occur in phases, with an initial upgrade to PandaX-20T planned to begin operation in 2027, before ultimately reaching a target mass of 43 tons in later phases~\cite{pandax_collaboration, PANDA-X:2024dlo}. 

This strategy builds on our proposal to search for annual modulation in the PMT dark rate of neutrino observatories~\cite{Leane:2025efj,Santos-Olmsted:2025nuk}. Xenon TPCs are an especially favorable realization, as their dark-electron rates are far smaller than the total PMT dark rates of large neutrino detectors, and their cryogenic, well-controlled environments allow excellent background control~\cite{LZ_WS2025_LowMassWIMP}. Furthermore, the large electroluminescence gain in dual-phase xenon TPCs enables better detection of weak ionization signals, which are typically more difficult to detect in neutrino experiments. Despite much smaller target masses, xenon TPCs can therefore provide stronger and more robust sensitivity. In our analysis using xenon TPCs, counting raw PMT hits rather than extracted electrons would not help further. Each electron already yields an S2 of $\sim$$30$ detected photoelectrons, so individual PMT hits add only the uncorrelated PMT dark rate ($\sim$$40$~Hz per PMT~\cite{Fruth:2019jmr,XENON:2024wpa}, or $\sim$$20{,}000$~Hz for the $\sim$$500$ PMTs in LZ and XENONnT) and photon-counting fluctuations, without adding signal.

\subsection{Feasibility of Signal Extraction}
\label{sec:feasible}

Annually modulating backgrounds can complicate a DM modulation search. However, Ref.~\cite{Leane:2025efj} showed with a Fisher analysis that, after profiling over the background modulation amplitude, the statistics-dominated scaling of the DM sensitivity ($\propto 1/\sqrt{B}$) is retained provided the modulation is sufficiently well constrained (see Ref.~\cite{Blanco:2026kda} for a related discussion of daily modulation). In practice, several sources modulate simultaneously, such as muon-induced activity, radon, solar neutrinos, and environmental effects on the detector, so separating them from the DM requires auxiliary information on their modulation~\cite{Leane:2025efj}. Importantly, the total background rate need not be known, as its normalization drops out of the fit; only the background modulation \textit{in phase} with the DM modulation must be constrained, at the level of $\sim$$5\times10^{-5}$ ($2\times10^{-4}$) of the total rate for one year of data at 50~Hz (5~Hz). While this is a significant task, it may be tractable for two reasons. First, no known background peaks at the DM phase; the closest is the muon flux, which peaks about a month later and whose modulation is measured directly by the muon vetoes. Second, modern xenon TPCs offer exquisite single-electron sensitivity in a stable cryogenic environment, and the dominant variations are tied to quantities they already record alongside the dark-electron rate. This includes the time and transverse position of every extracted electron, the larger S2 depositions that precede them, and the detector parameters that govern the background. Delayed electron emission in particular tracks the S2 history, so its response can be calibrated from the data themselves on short timescales rather than assumed stable. This auxiliary information can be used to constrain the background modulation without removing exposure, as we now discuss for the main components of the dark-electron rate.

The largest background component is S2-correlated delayed emission, which, outside of electrode hot spots, dominates the single-electron rate~\cite{XENON:2021qze,LUX-ZEPLIN:2025uaf,XENON:2024znc}. Because it follows the parent S2 deposition with a measured power-law time dependence~\cite{LUX-ZEPLIN:2025uaf}, this component can be modeled from the recorded S2 history together with continuously monitored detector parameters, such as the electron lifetime and extraction field. This modeling also removes the excess variance from the clustering of delayed electrons, in addition to constraining their modulation. The remaining uncorrelated component can be studied in a spatial sideband far from recent S2 depositions, as in Ref.~\cite{LUX-ZEPLIN:2025uaf}, and used to constrain its annual modulation directly. Finally, localized electrode emission, which can be episodic, is reduced with spatial and coincident-photon selections~\cite{LUX-ZEPLIN:2025uaf} at small cost to exposure.

Beyond these components, slow instrumental drifts, such as electrode conditioning, decreasing radon and krypton levels, and changes in purity, can leak into an annual fit, particularly with a single year of data. In the XENON100 keV-scale modulation searches, for example, a one-year annual excess did not persist with four years of data, where the data instead preferred a longer period and similar features appeared in control samples~\cite{XENON100:2015hgt,XENON:2017nik}. The drift-related quantities are continuously monitored and can be included in the fit, with their effect on the rate measured on short timescales. Unlike instrumental effects, a DM signal has a fixed annual phase that repeats every year, is uniform throughout the detector, and is independent of the time since the last S2 deposition. Multi-year data and the strong spatial and temporal variation of the background therefore provide further handles to separate it from a DM signal, without the exposure cuts of current searches.

A further asset is that three multi-ton xenon experiments are already operating --- XENONnT at LNGS, LZ at SURF, and PandaX-4T at CJPL --- and others are upcoming. They differ in depth, muon flux, radon environment, drift and extraction fields, purity, DAQ, and operating history, so their instrumental modulations are largely independent, whereas a DM signal should appear consistently in all of them with the same phase and with amplitudes fixed by their relative exposures and efficiencies. No single-detector analysis can provide a comparable test, and the data needed to perform it are already being recorded.

Importantly, we emphasize that the background modulation does not need to be controlled as well as our estimates assume (i.e., statistics limited) for this search to have leading sensitivity. Our estimates are stronger than the current cut-based xenon search~\cite{XENON:2024znc} by one to two orders of magnitude at low masses, so the in-phase background modulation can be larger than assumed by a comparable factor before that advantage is lost. In addition, because the sensitivity weakens only in proportion to the residual modulation, the analysis degrades gradually rather than failing; a residual a few times larger than assumed weakens the estimated sensitivity by only a factor of a few. A systematics-limited dark-electron rate analysis therefore remains more sensitive than the cut-based xenon approach unless the in-phase residual exceeds $\sim$$10^{-3}$ of the total rate at 50~Hz, or $\sim$$10^{-2}$ of the total rate at 5~Hz. It can also lead relative to the strongest limits from all other experiments~\cite{SENSEI:2020dpa,DAMIC-M:2025luv}, though by a smaller margin. These values are for a xenon target of about 6~tons; because the signal scales with target mass, the tolerable fraction is correspondingly larger for PandaX-xT and XLZD at the same rate.


\section{Conclusions and Outlook}
The single-electron regime is where most of the signal from the lightest DM is expected, yet it has been largely set aside because of large instrumental backgrounds. Only XENON1T and XENONnT have searched down to this regime, and to do so they applied severe cuts so that just $\sim$$10^{-4}$ of the exposure was retained. We have shown that this trade-off can be unfavorable for an annual-modulation search. Because the sensitivity then scales as $S/\sqrt{B}$ rather than $S/B$, retaining the bulk of the exposure can outweigh the larger background. We therefore propose using the bulk of the dark-electron rate itself, the raw electron hit rate previously generally treated as a nuisance, as the DM observable. Using measured dark-electron rates, we estimate the sensitivity of XENONnT, LZ, and PandaX-4T, and estimate the reach of PandaX-xT and XLZD. At the current rate of $\sim$50~Hz, these estimates improve on the cut-based xenon search by one to two orders of magnitude at low masses, and probe the parameter space of thermal freeze-out and freeze-in benchmark models. If the rate can be reduced to $\sim$5~Hz, the reach extends even further.

This strategy does not require including every raw electron. Rather, we emphasize that analyzing the continuously recorded dark-electron stream, instead of applying extensive exposure cuts, can substantially enhance sensitivity. Standard data-cleaning cuts, together with minimal cuts targeting the most intense backgrounds, are likely to be beneficial. An example of the latter is removing the brief ms window around large S2 depositions, which would reduce the exposure by only $\sim$$1\%$ or less. We leave a detailed optimization of the analysis to the experimental collaborations.

The analysis can begin now. It requires no new hardware. Triggerless DAQs, as implemented in XENONnT and PandaX-4T, already record the dark-electron rate throughout essentially the full runtime, and LZ's triggered DAQ records it within trigger windows. Our results therefore provide strong motivation for triggerless acquisition in current and future dual-phase TPCs. Moreover, three multi-ton xenon experiments at different sites are already recording these data, with more to come. A DM signal must appear in all of them with a common phase and the detector-specific amplitude, providing a powerful cross-check.

Several measurements would sharpen these estimates, including the dark-electron rate as a time series over multiple years, the stability of the delayed-emission response, and the size and seasonal variation of any component independent of the S2 history. Better understanding of delayed electron emission would also reduce single-electron backgrounds without large losses of exposure, directly improving the reach of this search. Finally, the same strategy can be applied to DM-nucleon scattering via the Migdal effect, which we will present in upcoming work.


\section*{Acknowledgments}
We are grateful for very helpful discussions with Dan Akerib, Tomasz Biesiadzinski, Tom Shutt, Ann Wang, and the broader LZ group at SLAC. We also thank Dan Akerib, Tomasz Biesiadzinski, Jianglai Liu, Ann Wang, and Ning Zhou for helpful comments on the draft. LSO and RKL are supported by the U.S. Department of Energy under Contract DE-AC02-76SF00515. JFB is supported by National Science Foundation Grant No.\ PHY-2609870 and was supported in part by the World Research Hub
(WRH) Program of Institute of Science Tokyo.


\medskip


\bibliography{main}
\bibliographystyle{apsrev4-2}

\end{document}